\documentclass{kapproc} 
\usepackage{t1enc}
\usepackage{procps} 
\usepackage[dvips]{graphicx}
\upperandlowercase
\let\footnote\savefootnote
\let\footnoterule\savefootnoterule
\kluwerbib 
\usepackage{latexsym}

\def\cf{cf.~}
\def\eg{e.g.~}
\def\eq{\!=\!}
\def\ltsim{~\rlap{\lower -0.5ex\hbox{$<$}}{\lower 0.5ex\hbox{$\sim\,$}}}
\def\gtsim{~\rlap{\lower -0.5ex\hbox{$>$}}{\lower 0.5ex\hbox{$\sim\,$}}}

\def\rbr{\hbox{$R_{\rm br}$}\,}
\def\rtr{\hbox{$R_{\rm tr}$}\,}
\def\hin{\hbox{$h_{\rm in}$}\,}
\def\hout{\hbox{$h_{\rm out}$}\,}
\def\rtrdh{\hbox{\rtr$/h$}}

\def\rbrdhin{\hbox{\rbr$/$\hin}}

\def\s0{S0}

\def\halpha{$\mathrm{H}$\@{\sc$\alpha$}\,}
\def\arcsec{\hbox{$^{\prime\prime}$}}
\begin{document}
\articletitle[Stellar Disk Truncations]
{Stellar Disk Truncations: \\ Where do we stand?}
\author{
M.~Pohlen\altaffilmark{1}, J.~E.~Beckman\altaffilmark{1}, 
S.~H\"uttemeister\altaffilmark{2}, J.~H.~Knapen\altaffilmark{3}, \\
P.~Erwin\altaffilmark{1}, and R.-J.~Dettmar\altaffilmark{2} 
} 
\affil{ 
\altaffilmark{1} Instituto de Astrof\'{\i}sica de Canarias  \\
\altaffilmark{2} Astronomisches Institut, Ruhr-Universit\"at Bochum \\
\altaffilmark{3} Centre for Astrophysics Research, University of Hertfordshire
} 
%
\begin{abstract}
In the light of several recent developments we revisit the phenomenon
of galactic stellar disk truncations. Even 25 years since the first
paper on outer breaks in the radial light profiles of spiral galaxies, 
their origin is still unclear. 
The two most promising explanations are that these 'outer edges' either 
trace the maximum angular momentum during the galaxy formation epoch,
or are associated with global star formation thresholds. 
Depending on their true physical nature, these outer edges may represent 
an improved size characteristic (e.g., as compared to $D_{25}$) and might 
contain fossil evidence imprinted by the galaxy formation and 
evolutionary history. 
We will address several observational aspects of disk truncations: 
their existence, 
not only in normal HSB galaxies, but also in LSB and even dwarf galaxies;
their detailed shape, not sharp cut-offs as thought before, but in fact 
demarcating the start of a region with a steeper exponential distribution of 
starlight; their possible association with bars; as well as problems 
related to the line-of-sight integration for edge-on galaxies (the main 
targets for truncation searches so far).
Taken together, these observations currently favour the star-formation 
threshold model, but more work is necessary to implement the truncations 
as adequate parameters characterising galactic disks.  
\end{abstract}
\section{Introduction}
The structure of galactic disks is of fundamental importance for
observationally addressing the formation and evolution of spiral 
galaxies. 
By measuring scalelengths for galaxy samples at different redshifts, it 
is for example possible to directly address the evolution of galactic disks
using the scalelength as a tracer of their sizes (\eg Labb{\' e} et al.~2003). 
To a first approximation, the radial light distribution of disks is well 
described by an exponential decline (Freeman 1970).  
However, for 25 years (since van der Kruit 1979), we have know that this
exponential light distribution does not extend to arbitrarily large radii.
Early, deep imaging using photographic plates showed that the disks are 
radially truncated at the outer parts (Fig.\ref{trunc}).
While truncations were difficult to measure at that time -- still true 
today for deriving their exact shapes -- it has become progressively 
easier to detect these {\it outer edges} with modern CCD equipment.  
In the optical they occur at galactocentric distances of roughly 2-6
exponential scalelengths (\cf next Sect.).  
Truncations could be used as an additional, intrinsic parameter 
measuring the total size of galactic disks (maybe replacing the 
widely used, but purely arbitrary $D_{25}$ definition), and truncations 
may be linked to the formation and evolution of disk galaxies in general.  
Despite the fact that they have been known for so long, we do not have 
a secure understanding of the origin for these outer edges, so their 
application as a fundamental disk characteristic is not yet possible.
In this overview we will summarise what is currently known about 
truncations and describe the quest to unveil their true origin. 
%
\section{Where are normal galaxies truncated?}
To understand their origin and use them as characteristic galaxy 
parameters (\eg in comparative studies of different galaxy samples), 
one has to empirically establish {\it where} `normal disks' are typically 
truncated -- conventionally expressed in terms of radial scalelength -- 
and {\it what} this truncation looks like. 
%
%
\subsection{The edge-on view}
\label{truncated}
\begin{figure}[t]
\hspace*{-1.0cm}
\includegraphics[angle=270,width=6.8cm]{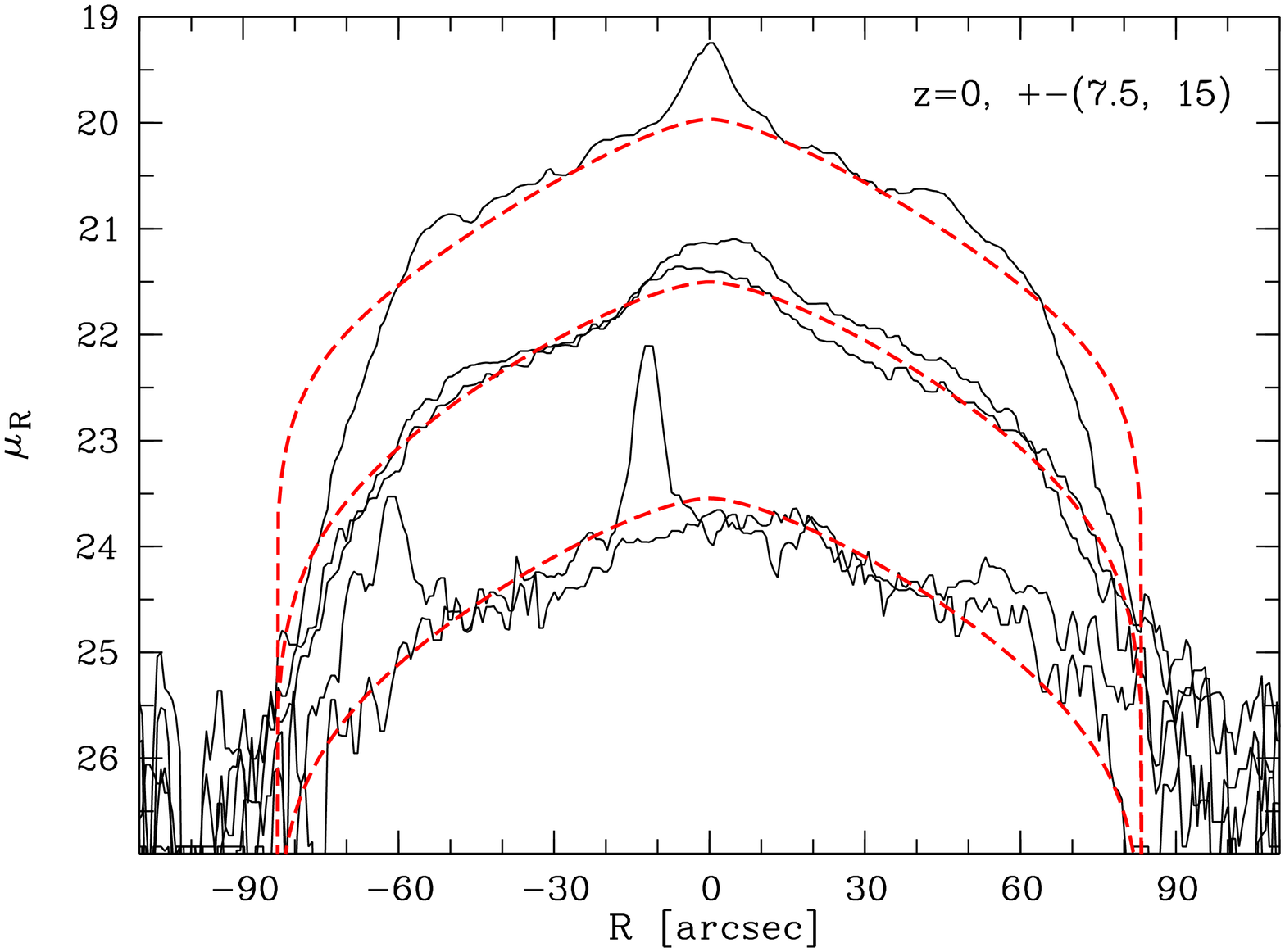}
\includegraphics[angle=270,width=6.8cm]{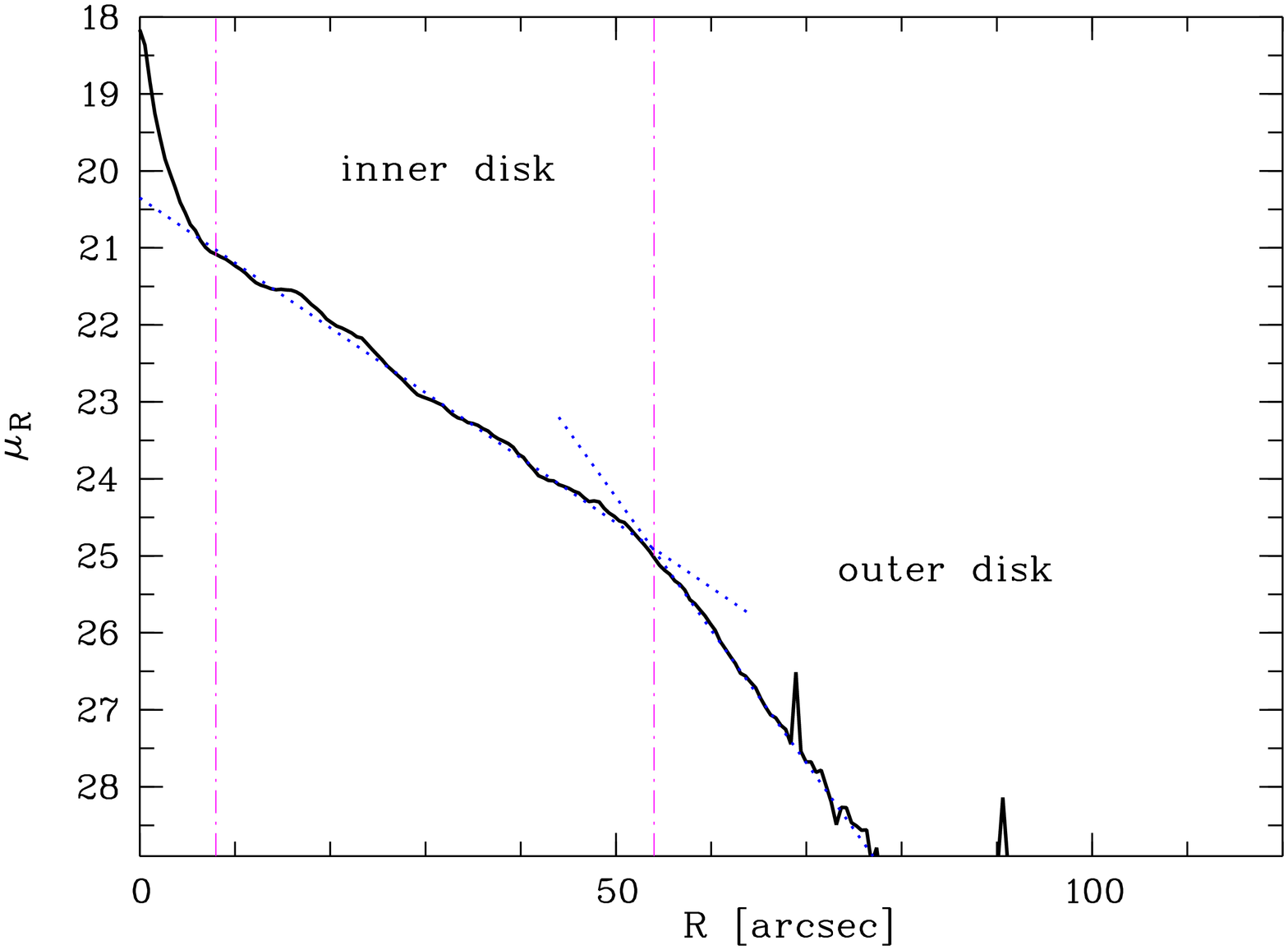}
\caption{ {\sl(a) Left panel:} Edge-on view: Radial surface brightness 
profiles {\sl (solid lines)} of the edge-on galaxy NGC\,522 together with 
the best fitting sharply-truncated model {\sl (dashed line)} \newline
{\sl(b) Right panel:} Face-on view: The azimuthally averaged radial surface 
brightness profile of the face-on galaxy NGC\,5923 {\sl (solid line)}. 
In addition to the inner part ($<8$\,arcsec) dominated by the bulge component 
an inner exponential slope {\sl (inner dotted line)} is clearly visible 
out to the break radius at $R_{\rm br}\!=\!54$\arcsec 
followed by another outer exponential part {\sl (outer dotted line)}. 
\label{trunc}
}
\vspace*{-0.0cm}
\end{figure}

In their seminal series of papers on the 3-D light distribution in 
edge-on galaxies, van der Kruit \& Searle (\eg 1981a, 1981b) measured 
the ratio of truncation radius ($R_{tr}$) to radial 
scalelength ($h$) as $4.2\pm0.6$ for a sample of 
seven nearby, large galaxies using photographic plates. 
They found that stellar disks have ``rather sharp edges'', justifying 
the terms {\it cut-off} and {\sl truncation} they used to describe them. 
\cite{bartel1994} collected more sensitive CCD images for a much 
larger sample of 27 edge-on galaxies -- later reanalysed and 
slightly enlarged by \cite{pohlen2000a} -- finding significantly 
lower ratios of $\rtrdh$, down to $2.9\pm0.7$ in the latter study. 
The Groningen group followed with their CCD survey providing 
a mean ratio of $3.8\pm1.0$ for a pilot sample of four galaxies
(de Grijs et al.~2001).
This lies in between the former two extremes, but clearly confirms
the explicitly larger scatter observed by \cite{pohlen2000a} compared 
to \cite{vdk81b}. 
Since line-of-sight (LOS) integration makes edge-on galaxies
ideal candidates to search for truncations, \cite{pohlen2001} 
obtained new data, trying to finally settle 
the true nature of the truncations by using significantly deeper exposures and 
selecting objects as undisturbed as possible. 
This study showed without doubt that the previously used concept of a 
sharp truncation does not really hold, as already suggested in
\cite{degrijs2001}.  
It is just possible to fit a sharply truncated model to 
this new data set (\cf Fig.\ref{trunc}a): \cite{pohlen2001} 
derived a value of $\rtrdh\eq 3.5\pm0.8$ using 56 galaxies, 
which is similar to the values of $3.6\pm0.6$ from the full analysis 
of 20 galaxies from the Groningen CCD survey (Kregel et al.~2002). 
However, it has become clear that the radial profiles are far better 
described by a {\it two-slope}, or {\sl broken exponential} structure,
characterised by inner and outer scalelengths and a well-defined 
break radius.
The deepest data show that LOS integration
-- a fairy godmother for detecting truncations -- is at the same 
time the major drawback in analysing them in detail (Pohlen 2001). 
Since disks are only empirically described by an exponential -- typically 
over at most a few scalelengths -- the edge-on geometry of such a two-slope 
disk, with additional dust obscuring the major axis, complicates an already 
delicate fitting process. 
Edge-on, there are always uncertainties in where to mark the break radius 
(\cf Kregel et al.~2002), and in particular the calculation of the true, 
unprojected scalelength is heavily model dependent. 
Within a sharply truncated model used by \eg \cite{pohlen2000b} there 
is a strong coupling between the two parameters \rtr and $h$.
However, fitting an infinite exponential model -- excluding the truncation 
region -- does not really solve the problem, especially for galaxies which 
exhibit an early break (in terms of scalelength) and a steep outer decline. 
The LOS integration will distribute the measured light along the assumed 
uniformly exponential decline so the best determined fitting scalelength 
will be systematically too small (\eg Pohlen et al.~2004).
A simpler approach is that of \cite{pohlen2001}, who avoided 3D, LOS
issues by fitting the observed profile with a 1-D, two-slope model,
finding $\rbrdhin\eq2.5\pm0.8$ from 37 galaxies. 
However, the fitted scalelengths in this case are not necessarily the 
intrinsic scalelengths, both because dust extinction increases the 
measured scalelength and because the scalelength of a 1D fit is larger 
than the intrinsic scalelength by as much as 20\% (e.g., de Grijs 1997).
%
%
\subsection{The face-on view}
The only way to settle the question of where the breaks in the radial profiles 
really occur is to go back to face-on galaxies as done earlier 
by \cite{vdk1988}. 
Therefore, \cite{pohlen2002} took deep, carefully flatfielded exposures of 
three face-on galaxies. The face-on geometry is practically independent of 
possible LOS effects caused by an integration across the multicomponent 
disk and much less affected by dust.
These data quantitatively establish the smooth truncation behaviour of the 
radial surface brightness profiles, which is best described by a two-slope 
or broken exponential model, characterised by inner \hin and 
outer \hout exponential scalelengths separated at a relatively well 
defined break radius \rbr (\cf Fig.\ref{trunc}b). 
The mean value for the distance independent ratio of break radius to 
inner scalelength -- which marks the start of the truncation region --
is $\rbrdhin\!=\!3.9\pm0.7$  for the three galaxies. 
This value is significantly larger than the value for the 
edge-on sample, but this is expected given the 
systematic errors in the scalelength measurement in the edge-on 
case described above.   
It is slightly lower compared to $\rtrdh\eq4.5\pm1.0$ for the 16 
face-on galaxies obtained by \cite{vdk1988}.
This small shift could be due to the different methods used 
to derive the truncation -- only estimated from isophote maps by 
\cite{vdk1988} -- or the sample sizes and selections.
%
\subsection{The combined view}
In summary, the radial profiles show a two-slope or broken exponential 
behaviour with an inner and outer disk separated by a rather sharp break 
radius at $2.5 \ltsim \rbrdhin \ltsim 5.8$, taking the mean value from the 
three face-on galaxies of \cite{pohlen2002} and the scatter from  
the larger edge-on sample of \cite{pohlen2001}.  
However, it is clear that truncations -- seen in different data sets, using
different methods, and analysed by different groups -- are {\it real}, in
contrast to the recent claim of \cite{narayan2003}.
It is not a matter of ``questioning their very existence'', but rather
of determining the shape of the profiles beyond the break radius, which is
admittedly still difficult to measure and a ``puzzle'' to explain.
%
%
%
\section{Are all disk galaxies truncated? }
\label{untruncated}
The data of Pohlen (2001) suggest that a very high fraction 
--more than 79\%-- of mostly late-type edge-on disks are 
truncated.
The remaining galaxies are either still disturbed (despite the 
careful selection), show an additional major structure not accounted 
for in a single component disk -- such as outer envelopes or a strong bar --, 
or (in one case) have S/N that is too low.
\cite{kregel} (see also Kregel 2003) support a rather high 
frequency of radial truncation features for a similar late-type 
sample of 34 edge-on galaxies.
They find successful truncation fits for 20 galaxies (59\%), whereas 
most of the other galaxies are rejected for technical reasons 
such as superimposed foreground stars or low S/N. 
The latter is the main problem of this sample in respect to truncations 
yielding only lower limits of typically $\rtrdh \gtsim 4.5$ for the other 
galaxies and just two cases of $\rtrdh \gtsim 6.0$.   
However, just recently \cite{erwin2004}, analysing a large 
sample of intermediately inclined, early-type galaxies, found seven 
barred galaxies without measured truncations beyond six scalelengths. 
Although the edge-on samples certainly also contain barred 
galaxies, we may ask if only barred galaxies (sometimes) 
lack truncations. 
It is true that the detailed studies by \cite{barton} and 
\cite{weiner}, which showed disks without truncation 
extending to at least seven and ten scalelengths, respectively, 
dealt with two strongly barred face-on galaxies. 
But the large collection of surface brightness profiles for unbarred or 
only weakly barred Sb--Sc galaxies by Courteau (1996) shows several cases 
of profiles which extend to at least six scalelengths without truncation, 
and the sample of \cite{erwin2004} also includes one unbarred Sa galaxy 
(UGC 3580) with $\rtrdh \gtsim 5.8$.  
It is still puzzling why the large edge-on samples -- which have the advantage
of LOS integration -- are missing many of the ``untruncated'' galaxies (or 
galaxies with truncations beyond six scalelengths) seen in 
face-on samples.
From the few late-type, face-ons in \cite{pohlen2002} and the large 
collection of early-type, face-ons in \cite{erwin2004} we know that 
azimuthal
averaging does not play a significant role in smoothing out truncations. 
If untruncated disks really do exist, then models which attempt to explain 
truncations also need to explain why some disks are {\it not} truncated.
%
\section{What is the origin of truncations?}
Despite the recent success in disentangling where the disks are truly 
truncated, the nature of galactic disk truncations is far 
from understood and their place in theories of disk galaxy formation 
is still under debate. 
The proposed hypotheses span a rather wide range of possibilities, 
though there are clearly two favoured explanation: 
the {\it collapse model} and the {\it threshold model}.
Van der Kruit (1987) deduced a direct connection to the galaxy formation 
process describing the truncations as remnants from the early collapse. 
The truncation reflects here the maximum angular momentum of the 
protogalaxy, which is completely independent of the hierarchical 
merging history.  
On the other hand, \cite{kenni} following Toomre (1964) has suggested 
a dynamical critical star-formation threshold ($\Sigma_{{\rm c}}$)
-- recently addressed in a different version by \cite{schaye2004} --
for thin, rotating, isothermal gas disks. 
If this persists over sufficient time it should produce a visible 
turnover or truncation in the observed stellar luminosity profile at 
that radius. Here the truncation radius corresponds to the radial 
position where $\Sigma_{{\rm gas}}(R)\eq\Sigma_{{\rm c}}$. 
More recently, \cite{vdb} combined the collapse and threshold models. 
While there is a truncation of the cold gas which reflects the maximum 
specific angular momentum of the baryonic mass that has cooled, the 
truncation radius of the {\it stars} reflects the presence of a 
star-formation threshold. 
Alternatively, \cite{zhang2000} and \cite{ferg3} have presented evolutionary 
scenarios -- typified as viscous evolution models -- which show features 
surprisingly comparable to the observed two-slope structure.  
However, according to \cite{ferg3} the breaks just reflect the initial 
conditions in the gas and tend to be smeared out while the galaxy evolves 
with time. 
\cite{batta} have proposed a connection with large scale galactic magnetic 
fields. This approach seems controversial, since it also explains flat 
rotation curves without dark matter; nevertheless, it is able to explain 
\eg possible $\rtrdh$ variations with morphological type or rotational 
velocity. 
The formation of a truncated stellar disk by tidal interaction, although 
appealing, is in contradiction to the radial spreading shown by 
\cite{quinn} in their numerical N-body simulations of satellite 
mergers. 
However, we are still missing an up-to-date N-body+SPH {\sl (stars and gas)}
simulation addressing a possible connection between truncations and 
interaction.   
%
%
\section{Can observations decide yet?}
\begin{figure}[t]
\hspace*{-1.0cm}
\includegraphics[angle=270,width=6.8cm]{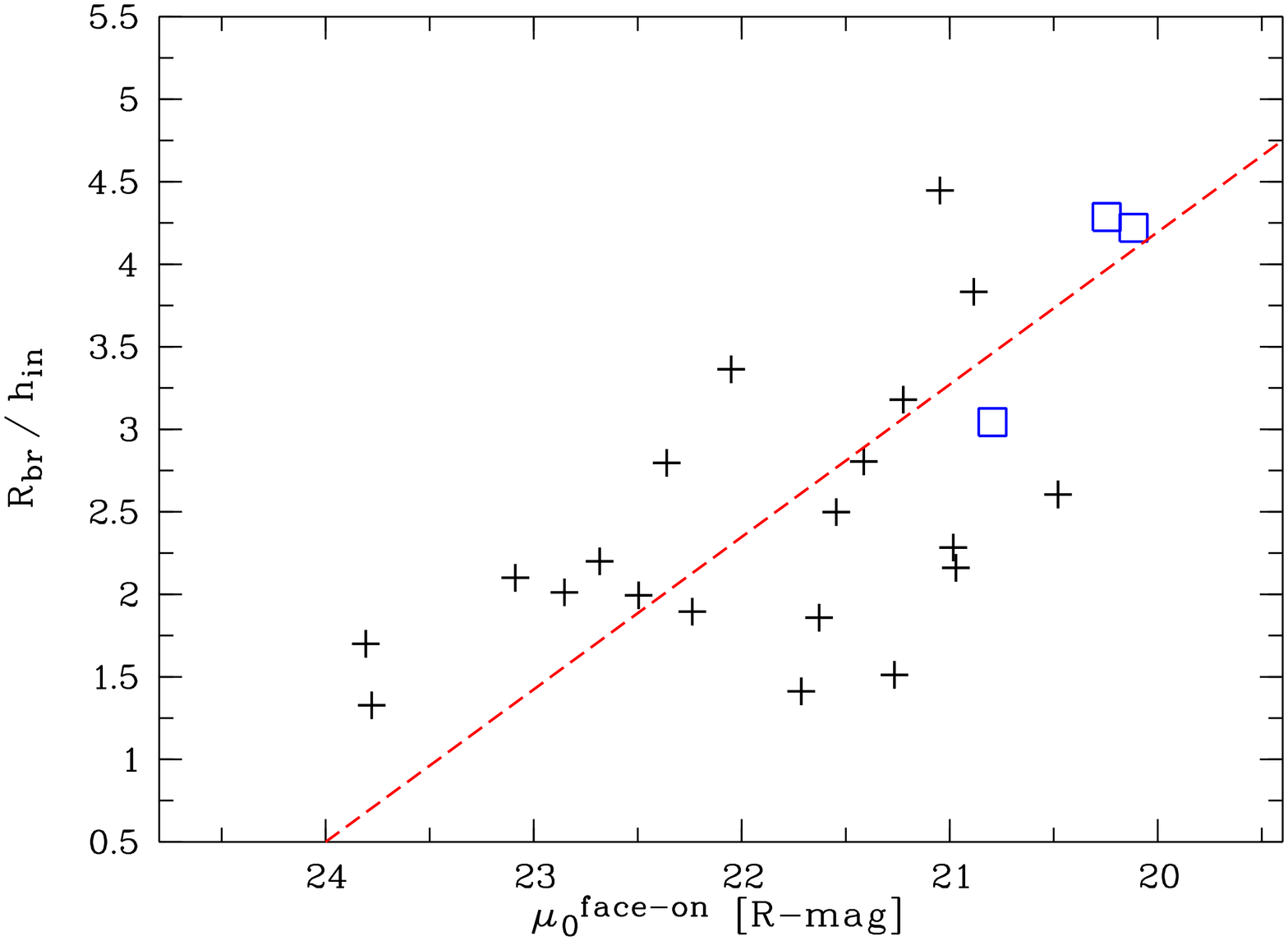}
\includegraphics[angle=270,width=6.8cm]{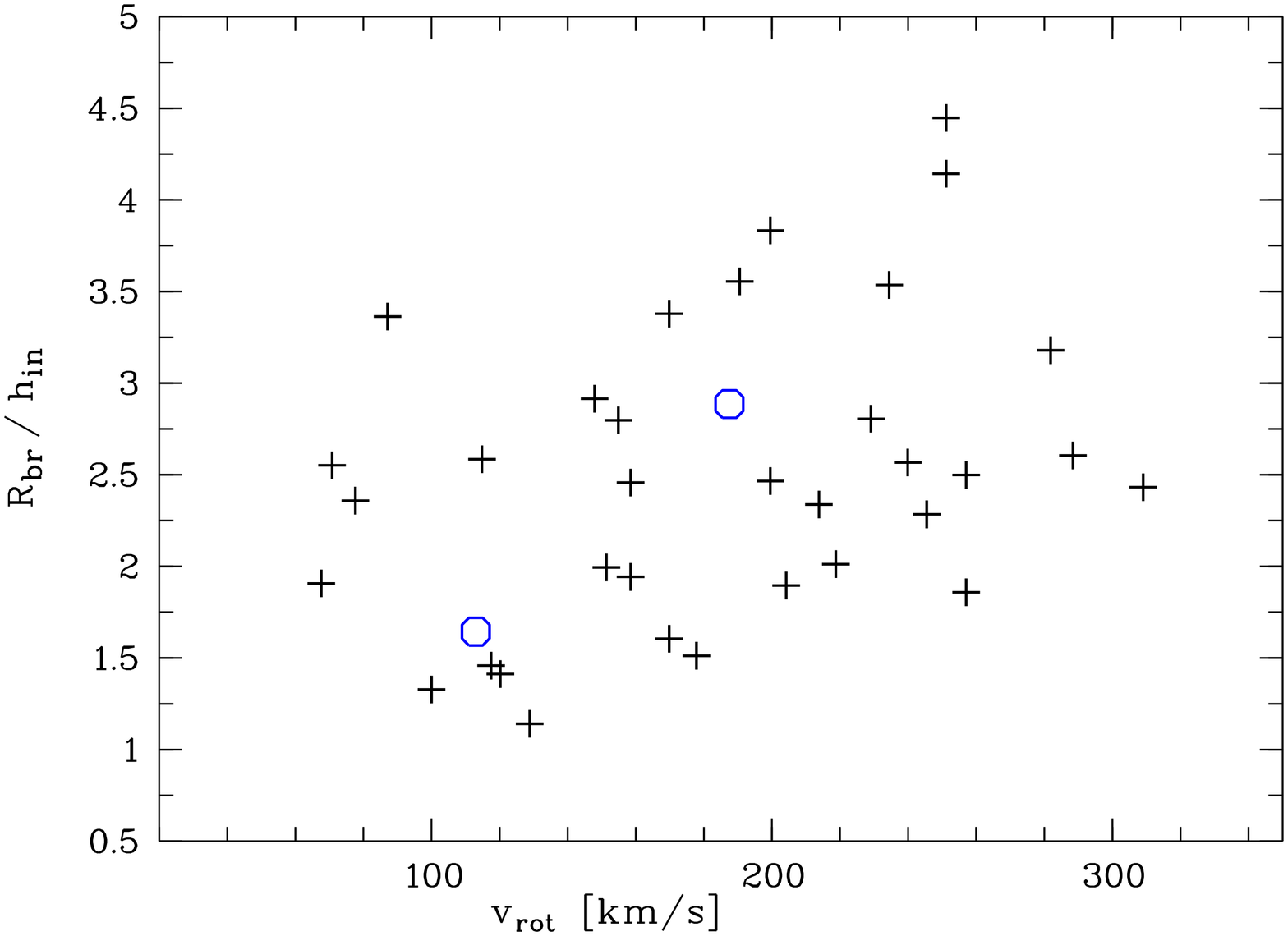}
\caption{ {\sl(a) Left panel:} Ratio of break radius to inner scalelength 
(\rbrdhin) versus face-on central surface brightness. {\sl Crosses:} 
edge-ons by \cite{pohlen2001}. {\sl Squares:} face-ons by \cite{pohlen2002}. 
{\sl Dashed line:} threshold prediction by \cite{schaye2004} according to 
\cite{kregelphd} \newline
{\sl(b) Right panel: }\ \rbrdhin versus maximum rotational velocity 
$v_{\rm rot}$.
{\sl Crosses:} edge-ons by \cite{pohlen2001}. {\sl Circles:} two face-on 
galaxies with known $v_{\rm rot}$ from the sample of \cite{pohlen2002} 
shifted to account for the mean systematic error of the two samples 
(\cf Sect.\ref{truncated}). 
\label{rtrdhvmu}
}
\end{figure}

\begin{figure}[t]
\hspace*{-1.0cm}
\includegraphics[angle=270,width=6.8cm]{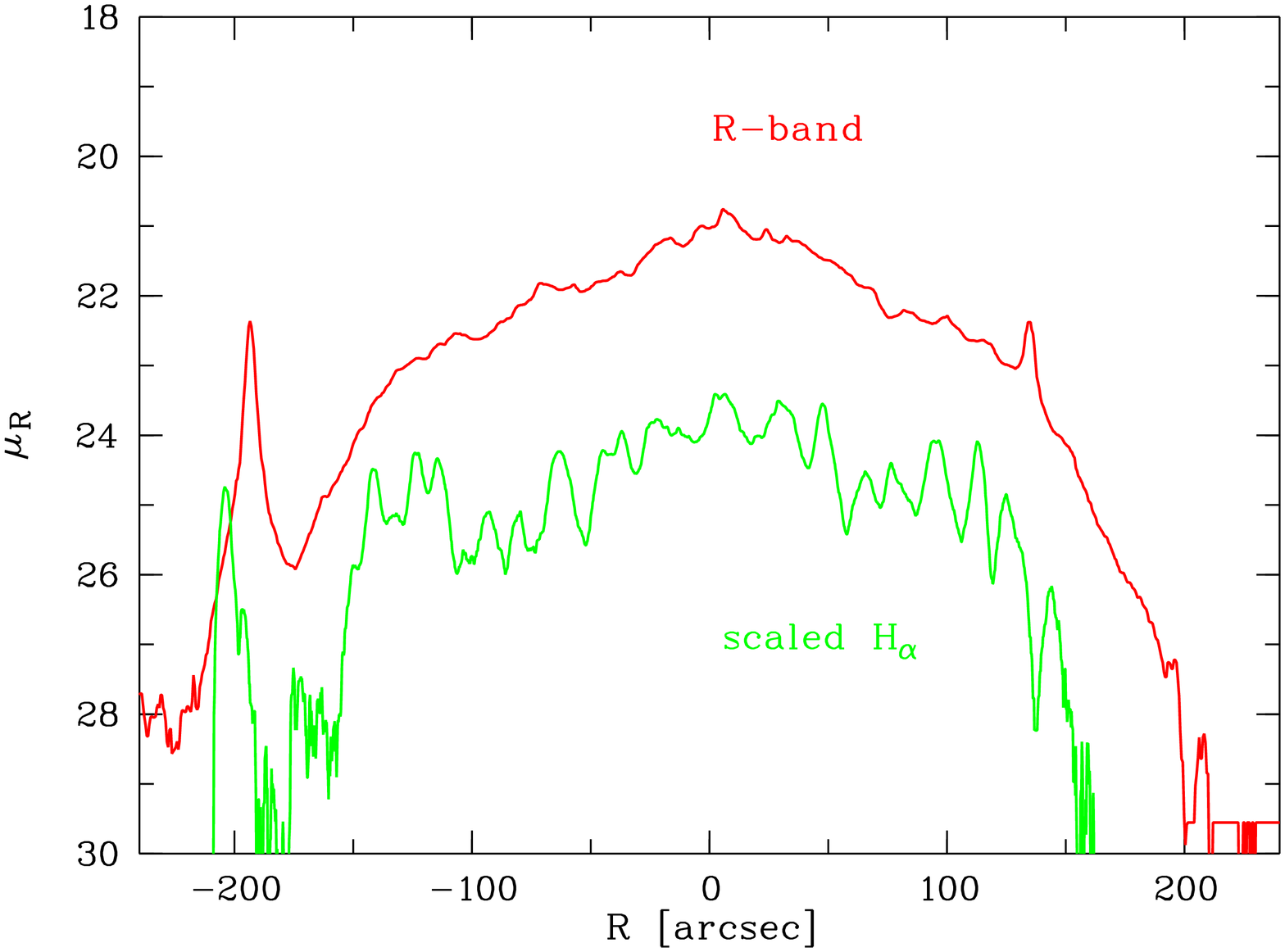}
\includegraphics[angle=270,width=6.8cm]{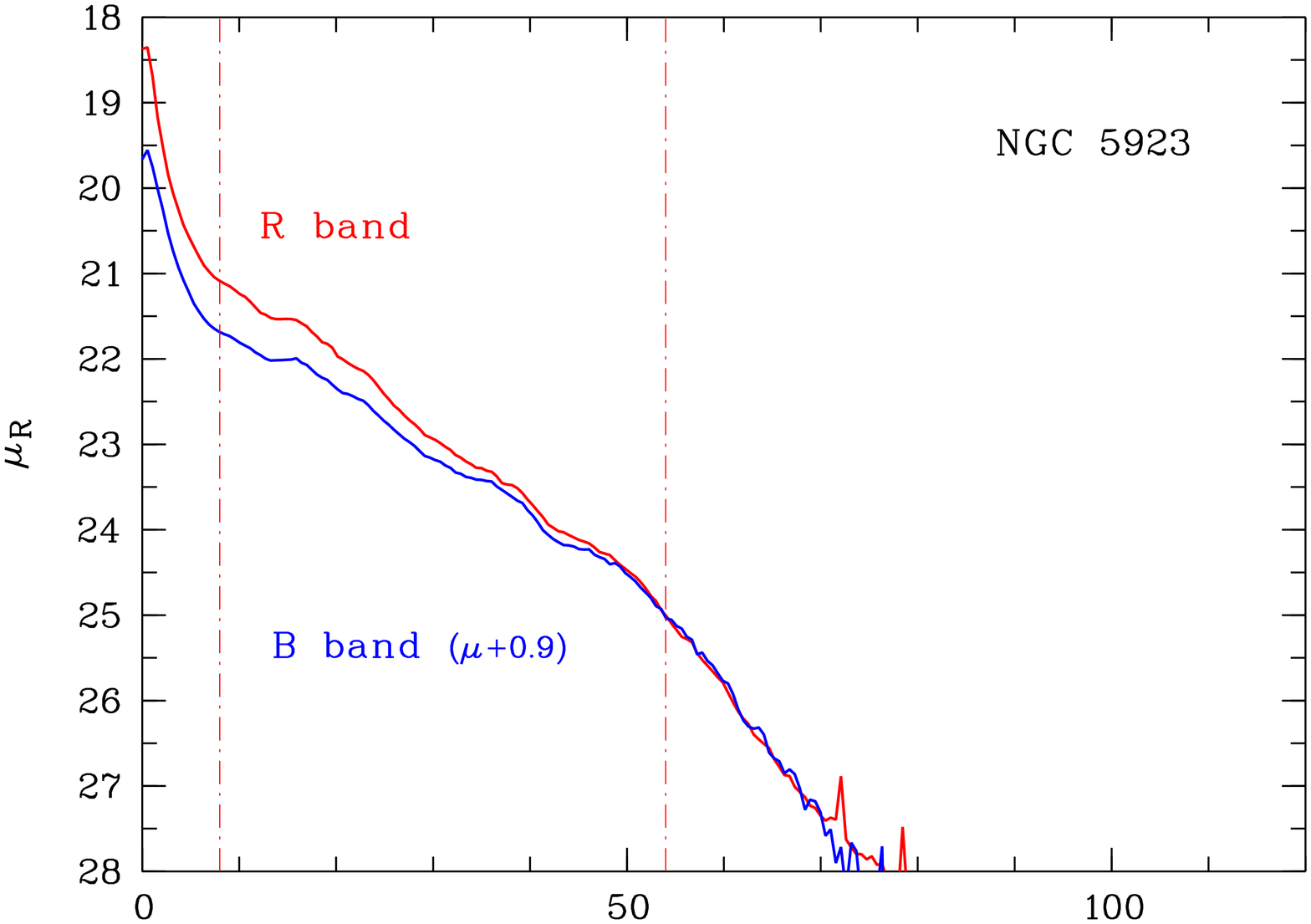}
\caption{ {\sl(a) Left panel:} Comparison of R-band major axis profile 
($6.2'$ vertically averaged) of UGC\,7321 and a scaled version of the 
continuum subtracted \halpha.
{\sl(b) Right panel:} The B- and R-band azimuthally averaged radial surface 
brightness profiles of the face-on galaxy NGC\,5923. \label{SFobs}
}
\vspace*{-0.0cm}
\end{figure}

The star-formation threshold models of \cite{schaye2004} and \cite{vdb} 
both predict a correlation of the ratio $\rtrdh$ with the central surface
brightness. Low-surface brightness galaxies should have lower values, as 
indeed indicated by current observations (\cf Fig.\ref{rtrdhvmu}a 
or Kregel 2003). 
The collapse model does not account for such a correlation since it predicts 
instead a fixed value for all galaxies, allowing for only very small scatter. 
According to the threshold models a similar trend should be observed in 
relation to the mass of the galaxy. 
Low-mass systems (characterised \eg by smaller rotational velocities) 
should have smaller values for the ratio of truncation radius to radial 
scalelength, as indicated by \cite{pohlen2001} (\cf Fig.\ref{rtrdhvmu}b). 
Additional support for a correlation with mass comes from recent 
work on nearby dwarf galaxies. \cite{hunter2002}, \cite{simon2003},  
and \cite{hidalgo} all find similar two-slope profiles with rather 
low values for $\rbrdhin$. 
%
%
The currently available large edge-on samples do not sufficiently cover 
the Hubble sequence to address the prediction by \cite{vdb} that 
early-type, bulge dominated galaxies (Sa to S0) should have $\rtrdh \ltsim 2$. 
However, in the large sample of moderately inclined, early-type, barred 
galaxies of \cite{erwin2004}, most galaxies do not follow this relation.
Only a small subset does, but these truncations are far better described as 
bar-related OLR breaks (\cf Sect.\ref{bars}). This failure of the promising 
threshold model of \cite{vdb} could be due to its rather general
implementation of the bulge formation process. 
%
%
\cite{kenni} (see also Martin \& Kennicutt 2001), measuring radial \halpha 
distributions, observed that the star-formation rate 
drops abruptly at a finite radius. 
Unfortunately only a few galaxies with both \halpha and optical breaks 
are currently available for testing the one-to-one correlation expected 
in a simple threshold scenario. 
Current data does indicate such a correlation 
(\cf Fig.\ref{SFobs}a or Hunter 2002), while beyond the break the 
decline in the \halpha luminosity seems to be more rapid than the 
stellar one.  
%
%
%
%
We know that disk galaxies have radial colour gradients: the disk 
gets bluer towards the edge (de Jong 1996). This implies that 
disks get radially younger -- arguing for an inside-out formation scenario 
-- in the sense that their inner regions are older and more metal rich than 
their outer regions (Bell \& de Jong 2000).  
However, this does not explain the truncation radius as the radius to 
which the disk has now grown (\eg Larson 1976), since the stars beyond
the break are indeed younger but still many Gyrs old. 
This is confirmed by \cite{davidge}, who studied the resolved stellar 
populations of NGC\,2403 and found that while young stars are restricted 
to the inner parts, AGB stars observed in the outer parts suggested 
that star-formation has occurred there during intermediate epochs of 1\,Gyr 
or more.  
The measured B- and R-band profiles of the face-on galaxy NGC\,5923 --
which trace the combined star-formation history -- indicate that the radial 
blueing extends only out to the optical break, which is at the same radius 
in both bands (\cf Fig.\ref{SFobs}b). 
Outside the break the outer disk has a constant blue colour 
$(B\!-\!R)\eq0.9$, which argues for either a single population or a coherent 
star-formation history.
NIR data may help to determine the question if the old stellar 
population -- tracing the mass of the galaxy -- changes across 
the break. 
Data by \cite{floridoNIR} show for the first time that truncations are 
also present in NIR images and that they are rather sharp. It turned out that 
these truncations are significantly smaller compared to the optical ones 
and it would be interesting to confirm these observations independently 
(\cf Pohlen 2001).
%
%
%
%
%
\section{Are bars an alternative solution? }
\label{bars}
Many barred galaxies possess rings in their outer disk produced by 
the bar's outer Lindblad resonance (OLR). 
While some of these rings are seen in radial profiles as mere bumps 
on an underlying exponential disk, \cite{erwin2004} found that a large 
fraction of their barred, early-type galaxies exhibit breaks  
associated with the OLR. 
Many of these {\sl OLR breaks} resemble the classical truncations 
described above, but located further inside. 
This implies that the bar is able to re-shape the initial gas disk 
(maybe replenished by later infall) out to substantial radii
($\approx\!1\!{\rm -}\!2\!\times\!\hin\!\approx\!2\!\times\!{\rm radius\ of\ bar}$).
Without additional data it is not yet clear if these OLR breaks 
also occur in more late-type galaxies -- known to have smaller 
and weaker bars -- thus providing a true alternative to the now 
favoured threshold theory (Erwin et al.~2004). 
However, the three clearly unbarred face-on Sbc--Sc galaxies 
of \cite{pohlen2002} argue against OLR breaks as the sole explanation 
for truncations, if one does not assume that these galaxies once had 
bars which are now dissolved. 
The intriguing existence of barred but untruncated galaxies 
(\cf Sect.\ref{untruncated}) implies that bars may also {\sl prevent}
truncations in some cases.
%
\section{Any open questions? }
The smooth nature of truncations clearly shows that there are stars 
outside the break radius. While the previously considered sharp truncations 
would agree nicely with the threshold model, a simple critical density does 
{\it not} account for the observed two-slope structure.
There are two possible explanations for these stars beyond the break
radius. 
They could be either born in situ -- out of an initial, maybe 
viscously redistributed, and already replenished gas disk -- 
or they are stars from the inner disk which have migrated 
outwards in a kind of diffusion process. 
\cite{ferg1998} already showed for a couple of galaxies that there 
is local star-formation at large galactocentric distances, probably 
beyond a break radius, supporting the former possibility.
However, both approaches still have to explain {\sl why} the inner and outer 
disk regions are produced with different, yet still exponential, slopes, 
{\sl why} the transition zone appears to be rather sharp, and -- even more 
challenging -- {\sl why} the outer disk is so large:
\cite{pohlen2002} showed for the face-on galaxies that the break radius 
is at \rbr$=\!\frac{2}{3}R_{\rm lmp}$ using the last measured point 
($R_{\rm lmp}$) as a reference.
In the case of the Milky-Way ($R_{\rm lmp} \ge 21$ kpc)
this implies an inner disk of 14\,kpc and an outer disk of at 
least 7\,kpc extra radius!
This is significantly larger than the usually assumed values 
of $\approx\!1$\,kpc previously deduced for a sharp truncation. 
According to P\'erez-Martin (this volume) there seems to be a 
trend of galaxies being truncated earlier (in terms of $h$)
at higher redshift (consistent with Tamm \& Tenjes 2003). 
So far, none of the described models has addressed such an 
interesting evolution with redshift. 
%
\section{Summary and Outlook}
Finally, we want to emphasise that the observed radial profiles 
show rather sharp break radii. However, they are not sharply 
truncated -- in the sense of an outer radius beyond which no further 
stars can be found -- as is often assumed in the literature. 
Despite the success of recent studies of the structural parameters, 
a medium-deep survey of a well selected sample of local, 
intermediate inclination to face-on disk galaxies along the Hubble 
sequence is still necessary, to provide ultimately {\it the reference
value} of $\rtrdh$.  
The star-formation threshold model seems to be the most promising 
hypothesis to explain the presence of truncations so far.
Although less direct than the collapse model -- which directly 
relates observed truncations to initial conditions -- van den Bosch's (2001)
model showed how to implement such a star-formation threshold in a collapse 
model which is able to trace the evolutionary picture. 
We are still working on additional observational support, such as 
more \halpha-to-optical break comparisons, detailed measurements of 
the atomic and molecular gas densities out to the break radius, 
and measurements of the stellar kinematics and population 
differences inside and outside the break.
%
%

\end{document}